# Superconductivity at 4.4K in PdTe$_2$-chains of a Ta Based Compound

Reena Goyal, Brajesh Tiwari, Rajveer Jha, and V. P. S Awana*

*CSIR-National Physical Laboratory, Dr. K. S. Krishnan Marg, New Delhi-110012, India*

**Abstract:**

Superconductivity in PdTe$_2$ heavy transition metal based compounds is rapidly developing field in condensed matter physics community. Here, we report superconductivity in a nominal ternary telluride Ta$_2$Pd$_{0.97}$Te$_6$ compound, which is synthesized in sealed evacuated quartz ampoule. The resultant compound is crystallized in layered monoclinic Ta$_4$Pd$_3$Te$_{16}$ phase with space group C2/m, having lattice parameters a=21.304(6)Å, b=3.7365(7)Å, and c=17.7330Å with β=120.65(1)°. Both transport and magnetic measurements demonstrated bulk superconductivity at 4.4K in studied polycrystalline sample. The metallic normal state conductivity can be well ascribed to Fermi-liquid behaviour. The superconducting upper critical field of the compound is determined to be 5.5Tesla based on magneto-restively measurements.

**\*Corresponding Author**
Dr. V. P. S. Awana, Principal Scientist
E-mail: awana@mail.npindia.org
Ph. +91-11-45609357, Fax-+91-11-45609310
Homepage www.fteewebs.com/vpsawana/

**Introduction:**

Searching new superconductors and understanding their superconducting mechanism and properties is one of the important driving forces for the condensed matter physics community. The discovery of high-temperature superconductivity in low-dimensional systems especially in Curates and Iron-based compounds is well celebrated and most haunted phenomenon in condensed matter physics in recent times [1, 2]. Recent observation of superconductivity in a strong spin-orbit coupled 4d and 5d transition metal based compound like $(Ta/Nb)_xPd(Te/Se/S)_y$ has been arousing high hopes for researchers due to their high critical field [3-9]. For a compound $Ta_2Pd_xS_5$ (x<1), Zhang *et al.* first reported type II superconductivity with $T_c$ up to 6K [3]. More recently, a new layered superconductor ternary telluride $Ta_4Pd_3Te_{16}$ superconductor with $T_c$ =4.5K at ambient pressure was found by Jiao *et al.* [9]. It is also claimed by Pan *et al.* that $Ta_4Pd_3Te_{16}$ is an unconventional superconductor of which $T_c$ first increases with pressure and later decreases above 3.1kbar applied pressure [10]. Here, we report bulk superconductivity at 4.4K in nominal $Ta_2Pd_{0.97}Te_6$ with its structure close to the $Ta_4Pd_3Te_{16}$ phase [9]. The identification of the structure of this type of compound by Rietveld refinement is based on the available reports of compound like $(Ta/Nb)_xPd(Te/S/Se)_y$. The superconducting properties of this compound are characterized by magneto-resistivity and magnetic susceptibility measurements. The superconducting upper critical field ($H_{c2}$) of the compound is determined to be 5.5Tesla.

**Experimental details:**

Polycrystalline bulk compound with nominal composition $Ta_2Pd_{0.97}Te_6$ was synthesized via solid state route. We mixed uncontaminated mixture of Ta, Pd and Te in a stoichiometry ratio of 2:0.97:6 in argon controlled glove box and then pelletized by applying uniaxial stress of 100kg/cm$^2$. The pellets were sealed in an evacuated (< 10$^{-3}$ Torr) quartz tube and kept in the furnace for heating at temperature 850$^0$C with a rate of 2$^0$/min for 24h. The same sample was again pulverised and pelletized before being heated for another 24h at same precondition. The structural characterisation was done with Rigaku x-ray diffractometer using Cu K$_\alpha$ line of 1.54184Å. Electrical and magnetic measurements were performed with the assistance of Quantum Design (QD) Physical Property Measurement System (PPMS) - 140 kOe down to 2K.

**Results and discussion**

The observed XRD pattern along with its Rietveld refined simulated pattern with in monoclinic C2/m (#12) structure using Full-Prof is shown in Figure 1(a) for nominal $Ta_2PdTe_6$. The global fitness of XRD pattern is $\chi^2 = 3.21$. The values of the lattice parameters are also given in Figure1. The XRD analysis reveals that the synthesized compound crystallizes in layered monoclinic structure with in space group C2/m. The schematic cell is shown in the right panel of Figure1. The obtained lattice parameters are a=21.304(6)Å, b=3.7365(7)Å, c=17.7330Å and β=120.65(1)°. Table 1 presents the Wyckoff positions of inequivalent atomic sites with their fractional coordinates. The refined chemical composition turns out to be $Ta_{1.3}Pd_{0.91}Te_{6.97}$, which is though not far away from the starting nominal composition, but is crystallized in $Ta_4Pd_3Te_{16}$ phase [9]. Ideally, one could expect that $Ta_2PdTe_6$ may crystallize in tetragonal $Nb_2PdS_5$ phase [3-7]. Interestingly, the $Ta_4Pd_3Te_{16}$ phase does not crystallize with same nominal composition, probably due to Pd/Te loses [9]. In our case, we lead to $Ta_4Pd_3Te_{16}$ phase with $Ta_2Pd_{0.97}Te_6$ nominal composition. Further, instead of single crystals, we have stabilized the structure in bulk polycrystalline form with same superconducting characteristics as in ref. 9. The crystal structure clearly shows a chain of Pd1 atoms along crystallographic b-axis with pyramidal structure of $Ta_2Te_3$ with Pd1 being base centred. This chain of Pd1 can play dominant role in electronic and superconducting properties as its bond length with Ta and Te are almost same. In a recent pre-print D. J. Singh proposed that $Ta_4Pd_3Te_{16}$ is a multiband superconductor with an electronic structure mostly derived from Te p states [11]. The role of this chained structure and hybridization of Pd with Ta and Te orbital may reveal more on its superconducting properties.

Resistivity measurement with the variation of temperature is shown in Figure 2, which clearly indicates metallic normal state with a superconducting onset $T_c^{onset}$ = 5.4K and $T_c$ ($\rho_0$) =4.4K. The resistivity obeys $\rho=\rho_0+AT^2$ (solid red line in Figure2) in temperature range of 6K to 35K, suggesting that dominant scattering at low temperatures is electron-electron type, a hallmark of Fermi-liquid. The residual resistivity ($\rho_0$) by extrapolation gives the value of 5.74x10$^{-7}$ Ω-cm. The residual resistivity ratio (RRR; the ratio of the resistivity at room temperature and at zero temperature) is estimated to be 6.23, which is though good enough for polycrystalline bulk sample but much smaller than the recently studied single crystal of $Ta_4Pd_3Te_{16}$ [9]. Lower inset of Figure 2 shows the magneto-resistivity being measured perpendicular to the applied magnetic field of up to 5Tesla. With increasing magnetic field, $T_c$ decreases at the rate of 0.6K/Tesla, see upper inset of Figure 2. Based on 90% criteria of normal state resistivity (90% $\rho_n$), the estimated upper critical field being determined using the WHH formula, i.e., $H_{c2}(0)=-0.69(dH_{c2}/dT)T_c$ at absolute zero temperature is 5.5Tesla, which is almost at the border of Pauli paramagnetic limit [12]. $H_{c2}(T)$ is understood as the field at which magnetic flux lines fill the entire volume of a sample.

Magnetic susceptibility ($4\pi\chi$) as shown in Figure 3 is recorded in zero field cooled (ZFC) and field cooled (FC) conditions under applied field of 10Oe as a function of temperature, which confirms strong diamagnetic property below 4.4K for nominal $Ta_2Pd_{0.97}Te_6$ and $Ta_4Pd_3Te_{16}$ phase crystallized compound. As clear from ZFC magnetic susceptibility, the superconducting volume fraction or diamagnetic shielding of the sample is though above 75%, the FC curve that presents the Meissonier expulsion of applied magnetic field (10Oe) is about 22% at 2K. The partial flux retention due to flux trapping is above 50% even for small field. Inset of Figure 3 shows the real $\chi'$ and the imaginary $\chi''$ parts of ac magnetic susceptibilities recorded at 333Hz frequency with 10Oe amplitude, while the bias field kept zero. It is well known that ZFC dc $\chi$ is equivalent to real part of ac $\chi'$, which can be easily verified by comparing [13]. An interesting difference one can observe that close to superconducting transition the dc magnetic susceptibility shows clear paramagnetic moment, which is not present in ac magnetic susceptibility. The peak in $\chi''$ is due to development of superconductivity as a filamentary network while allowing the magnetic flux to penetrate [13].

Figure 4 shows the isothermal magnetization curves recorded at 2K and 3K in the superconducting state. As magnetic field increases from zero, the magnetization decreases linearly up to $H_{c1}$ (=0.15kOe at 2K) suggesting diamagnetic character. Above the magnetic field $H_{c1}$, the magnetization starts increasing, reaches to zero and becomes positive above applied field of 4.2kOe. It is worth mentioning that at 2K, the resistivity is zero even in the applied field of 20kOe (see figure 2) whereas from the M-H curve, the magnetization become positive just above 4.2kOe, which is much lower suggesting inhomogeneity of superconducting fraction and large demagnetizing fields being present in the sample. Another justification for this anomaly can be presence of percolation path of superconductor in the bulk of superconductor, which makes resistivity zero even if the overall sample is not diamagnetic. The deviation of upper critical field by magneto-resistivity measurements and magnetization measurements needs a serious attention to rule out the contributions from demagnetizing fields and inhomogenities in heavy metal-based superconductors before attributes this behaviour to be intrinsic.

**Conclusion:**

In summary, we have studied the superconducting properties of very recently discovered low-dimensional compound $Ta_4Pd_3Te_{16}$ in bulk polycrystalline sample. XRD analysis indicated that nominally taken $Ta_2Pd_{0.97}Te_6$ composition is crystallized in layered $Ta_4Pd_3Te_{16}$ phase with monoclinic structure. Seemingly, the loss of Pd during heating requires the off stoichiometric starting composition. The metallic normal state at low temperature indicated Fermi-liquid nature of conduction electrons. Using 90% of normal resistivity criteria, the upper critical field estimated to be 5.5Tesla, which is well within Pauli-paramagnetic limit. This is the first report on synthesis of bulk polycrystalline $Ta_4Pd_3Te_{16}$ phase with $T_c$ of above 4.4K, and seconds the same as being reported recently for the single crystals of the this material [9].

**Acknowledgement:**

Authors would like to thank their Director NPL India for his interest in the present work. This research work is financially supported under *DAE-SRC* outstanding investigator award scheme on search for new superconductors.

**Figure caption:**

Figure 1. (a) (Color online) Room temperature Reitveld fitted powder XRD pattern of nominal $Ta_2Pd_{0.97}Te_6$ with in $Ta_4Pd_3Te_{16}$ phase, the difference (blue line) is minimized between observed (open red circle) and calculated (solid black line) patterns. The position of allowed Bragg reflections are shown as bars (pink). (b) Crystal structure of $Ta_4Pd_3Te_{16}$ phase with a clear chain of Pd1 atoms along crystallographic b-axis and pyramidal structure of $Ta_2Te_3$ with Pd at base centred.

Figure 2. Resistivity as a function of temperature from 300K to 2K is showing a clear superconducting transition at 4.4K. Low temperature resistivity well fitted to $\rho=\rho_0+AT^2$ (solid red line). Lower inset presents the magneto-resistivity at various applied magnetic fields of up to 5Tesla. The upper inset shows the H-T diagram as derived from the criteria 90% of normal state resistivity.

Figure 3. dc magnetic susceptibility ($4\pi\chi$) recorded in zero field cooled (ZFC) and field cooled (FC) conditions with 10Oe applied magnetic field. Inset presents ac magnetic susceptibilities; real $\chi'$ and imaginary $\chi''$ at 333Hz frequency and 10Oe amplitude.

Figure 4. Isothermal magnetization (M) in superconducting state (at 2K and 3K) of $Ta_2Pd_{0.97}Te_6$ as a function of applied magnetic field (H). Inset presents low field one coordinate M-H curve clearly marking the lower critical field ($H_{c1}$).

**Table. 1 Wyckoff positions, site symmetry, fractional coordinate and occupancy of different atoms.**

| Atom | Wyck. (sym) | x | y | z | Occupancy |
|------|-------------|----------|---|----------|-----------|
| Te1  | 4i(m)       | 0.316(2) | 0 | 0.073(2) | 1.00      |
| Te2  | 4i(m)       | 0.110(2) | 0 | 0.038(2) | 0.80      |
| Te3  | 4i(m)       | 0.284(3) | 0 | 0.748(2) | 0.90      |
| Te4  | 4i(m)       | 0.191(2) | 0 | 0.376(2) | 0.97      |
| Te5  | 4i(m)       | 0.580(2) | 0 | 0.171(2) | 0.80      |
| Te6  | 4i(m)       | 0.016(3) | 0 | 0.751(3) | 0.50      |
| Te7  | 4i(m)       | 0.400(3) | 0 | 0.363(2) | 0.50      |
| Te8  | 4i(m)       | 0.122(2) | 0 | 0.490(2) | 0.70      |
| Pd1  | 4i(m)       | 0.229(3) | 0 | 0.186(3) | 0.80      |
| Pd2  | 2c(2/m)     | 0        | 0 | 0.5      | 0.11      |
| Ta1  | 4i(m)       | 0.710(1) | 0 | 0.082(1) | 0.80      |
| Ta2  | 4i(m)       | 0.597(2) | 0 | 0.351(2) | 0.50      |

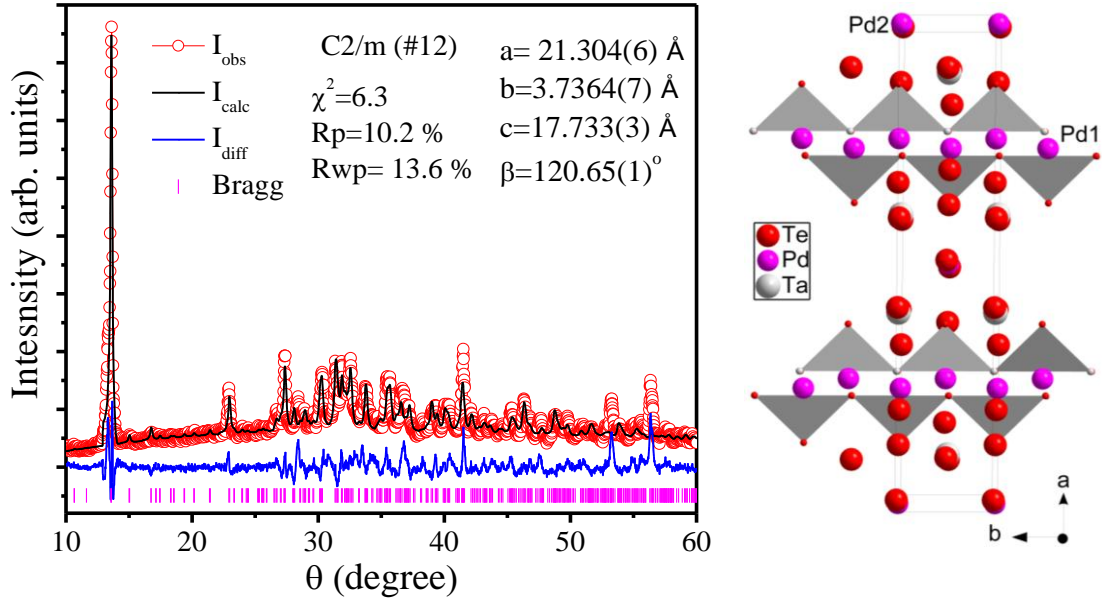

Figure 1

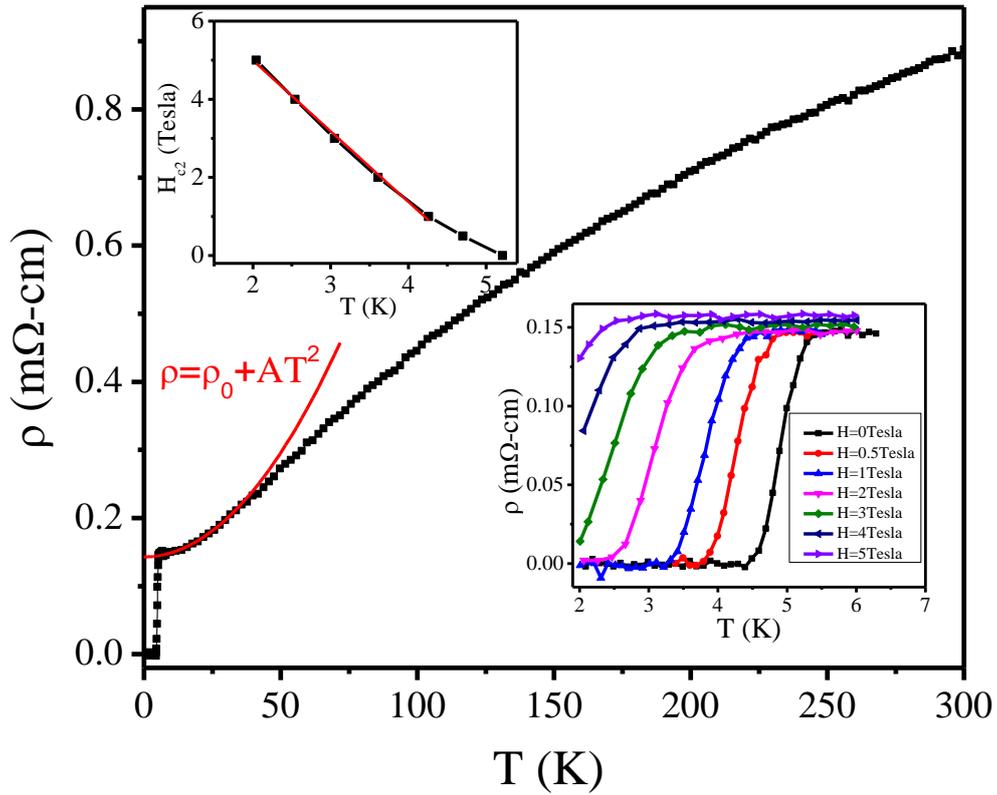

Figure2

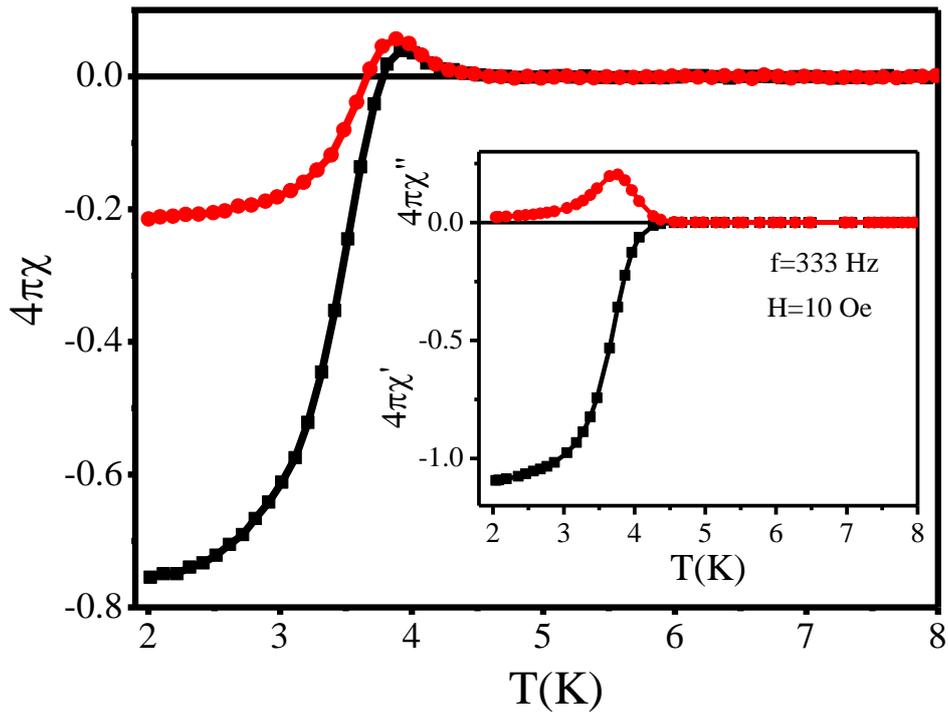

Figure3

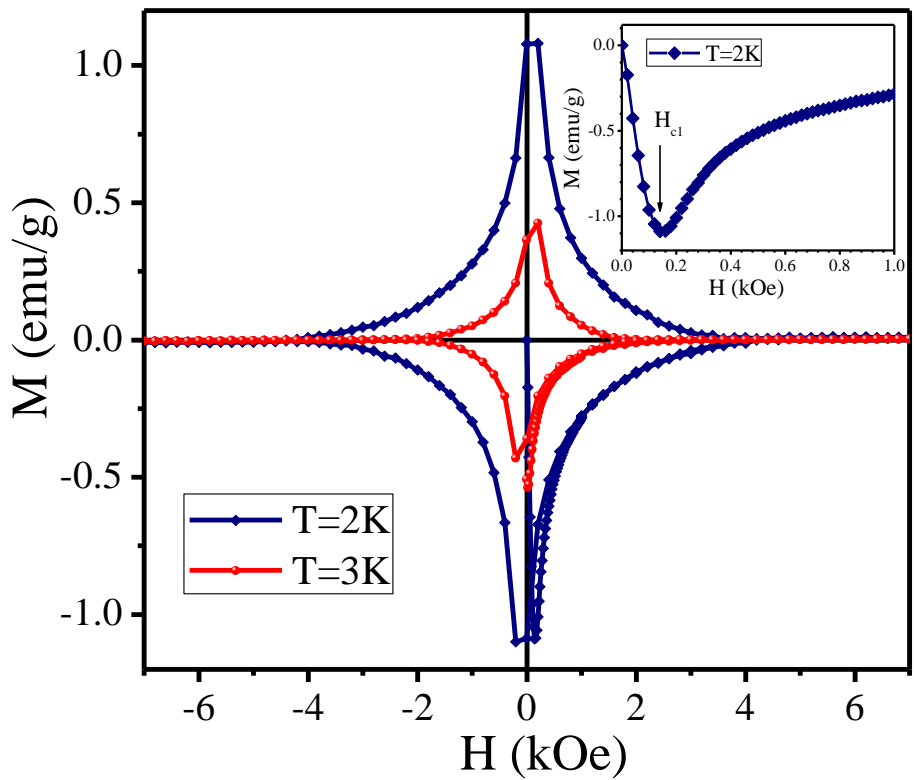

Figure4